\documentclass[twocolumn, superscriptaddress, prl,floatfix]{revtex4-1}
\usepackage{graphicx}
\usepackage[integrals]{wasysym} 
\usepackage{nicefrac,amssymb,amsmath} 
\usepackage{physics}
\usepackage{color}
\newcommand{\tise}{TiSe$_2$}

\newcommand{\bRcal}{\boldsymbol{\Rcal}}
\newcommand{\Rcal}{\mathcal{R}}
\newcommand{\bPhi}{\boldsymbol{\Phi}}

\newcommand{\bLambda}{\boldsymbol{\Lambda}}
\newcommand{\tcdw}{T$_{\textrm{CDW}}$}
\newcommand{\te}{T$_e$}

\newcommand{\hightt}{high-$T$}
\newcommand{\lowtt}{low-$T$}
\newcommand{\Dcal}{\mathcal{D}}
\newcommand{\bDcal}{\boldsymbol{\Dcal}}

\newcommand{\qe}{\texttt{QUANTUM ESPRESSO}}
\newcommand{\crystal}{\texttt{CRYSTAL}}

\begin{document}

\title{Anharmonic melting of the charge density wave in single-layer TiSe$_2$}

\newcommand{\Roma}{ Dipartimento di Fisica, Universit\`a di Roma Sapienza, Piazzale Aldo Moro 5, I-00185 Roma, Italy}
\newcommand{\Genova}{Graphene Labs, Fondazione Istituto Italiano di 
Tecnologia, Via Morego, I-16163 Genova, Italy}
\newcommand{\Spain}{Fisika Aplikatua 1 Saila, Gipuzkoako Ingeniaritza 
Eskola, University of the Basque Country (UPV/EHU),
Europa Plaza 1, 20018, Donostia-San Sebasti\'an, Basque Country, Spain}
\newcommand{\CFM}{Centro de F\'isica de Materiales (CSIC-UPV/EHU),
             Manuel de Lardizabal pasealekua 5, 20018 Donostia-San 
Sebasti\'an, Basque Country, Spain}
\newcommand{\Spainnn}{Donostia International Physics Center (DIPC),
             Manuel de Lardizabal pasealekua 4, 20018 Donostia-San 
Sebasti\'an, Basque Country, Spain}
\newcommand{\insp}{Sorbonne Universit\`e, CNRS, Institut des Nanosciences de Paris, UMR7588, F-75252, Paris, France}
\author{Jianqiang Sky Zhou}
\email[]{zhou@insp.jussieu.fr}
\affiliation{\insp}
%\affiliation{\etsf}
\author{Lorenzo Monacelli}
\affiliation{\Roma}

\author{Raffaello Bianco}
\affiliation{\CFM}

\author{Ion Errea}
\affiliation{\CFM} 
\affiliation{\Spain}
\affiliation{\Spainnn}
\author{Francesco Mauri}
\affiliation{\Roma}
\affiliation{\Genova}

\author{Matteo Calandra}
\email[]{matteo.calandra@upmc.fr}
\affiliation{\insp}

\begin{abstract}
Low dimensional systems with a vanishing band-gap and a large electron-hole interaction
have been proposed to be unstable towards exciton formation. 
As the exciton binding energy increases in low dimension,
conventional wisdom suggests that excitonic insulators should be more stable in 2D than in 3D.
 Here we study the effects of the electron-hole interaction and anharmonicity in single-layer TiSe$_2$. We find that, contrary to the bulk case and to the generally accepted picture, the electron-hole exchange interaction is much smaller in 2D than in 3D and it has negligible effects on phonon spectra. By calculating
anharmonic phonon spectra within the
stochastic self-consistent harmonic approximation,  
we obtain \tcdw$\approx440$K for an isolated
and undoped single-layer and \tcdw$\approx364$K for an electron-doping $n=4.6\times10^{13}$ cm$^{-2}$, close to the experimental result of $200-280$K on supported samples. Our work demonstrates
that anharmonicity and doping melt the charge density wave in single-layer TiSe$_2$.
\end{abstract}

\date{\today}%
\maketitle

\begin{figure}
    \centering 
    \includegraphics[width=1.0\linewidth]{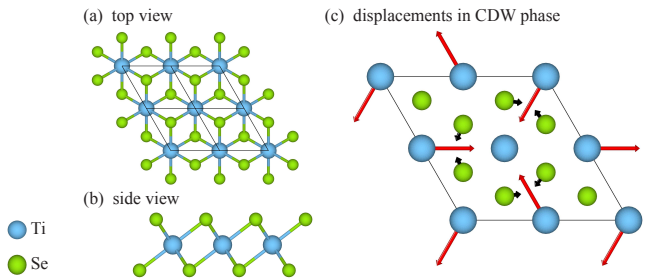}
    \caption{(a) and (b): top and side schematic views of crystal structure of monolayer \tise\ in the \hightt\ phase on a $2\times2$ cell. (c): top view of the atomic displacements of the \lowtt\ phase (i.e., the CDW phase) with respect to the \hightt\ phase. Blue and green balls represent Ti and Se atoms, respectively.}
    \label{fig:displacement}
\end{figure}

The occurrence of charge ordering in bulk \tise\ (see Fig. \ref{fig:displacement}) and its possible interplay with electronic excitations has attracted increasing interest over the last years. Two scenarios for the occurrence of the charge density wave (CDW) have been proposed: the first one is purely electronic and  is based on exciton condensation \cite{KohnGreyTin,KeldyshKozlov,exciton-insulator-1965,JeromePhysRev.158.462,Kogar2017-science}, while in the second the lattice plays a dominant role via the electron-phonon interaction \cite{Calandra-PRL-2011-EP-tise2,WeberPhysRevLett.107.266401,MCalandra-PRL2017}.
However, both scenarios are incomplete, as there are currently no explanations of the strong temperature dependence of phonon spectra in the \hightt\ \cite{WeberPhysRevLett.107.266401}
and \lowtt\ phases \cite{HolyPhysRevB.16.3628,SnowPhysRevLett.91.136402} and of the magnitude of \tcdw.
Surprisingly, little is
known about anharmonicity and its effect on the CDW in \tise .

From the theoretical point of view, it has been shown that harmonic calculations in bulk \tise\ including the electron-phonon
interaction within density-functional perturbation theory (DFPT) \cite{RevModPhys-2001-DFPT} correctly reproduce the occurrence of a CDW with a $2\times2\times2$ periodicity
 \cite{Calandra-PRL-2011-EP-tise2,MCalandra-PRL2017,Bianco-prb-2015}. 
However, the electronic structures of the high- and low-$T$ phases
as well as Raman and infrared spectra of the \lowtt\ phase at $T=0$K can only be explained by including the electron-hole exchange interaction within hybrid functionals \cite{MCalandra-PRL2017}. Density functional theory (DFT) with semi-local kernels leads to a metallic electronic structure, in disagreement with the angle-resolved photoemission spectroscopy (ARPES) experiments \cite{Calandra-PRL-2011-EP-tise2, Bianco-prb-2015,Rohwer2011-ARPES-bulk} that show a weakly doped semiconductor in both phases.
Moreover, they underestimate the square of the electron-phonon deformation potential of a factor of $3$ \cite{MCalandra-PRL2017}. 
%The anharmonic phonon calculation in bulk \tise\ requires then the inclusion of the exchange interaction and it is thus very cumbersome.

Recently, single-layer \tise\ was synthesized either by exfoliation or molecular beam epitaxy (MBE). It displays a $2\times2$ CDW with a \tcdw\ that is enhanced with respect to the bulk case (T$_{\rm CDW}^{\rm bulk}\approx 200 $K) and is strongly substrate dependent \cite{2DMaterials-2018-substrate-dependent-TC,Wang2018-advmat,Li2016-nature,acs-nano-2017-Duong,Sugawara-ACSNANO-2016,Chen2015-nature-commu,Fang-prb2017,Chen2015-nature-commu,Fang-prb2017}: single-layer \tise\ on top of insulating MoS$_2$ has \tcdw$=280$K \cite{2DMaterials-2018-substrate-dependent-TC}, 
while on top of $n$-doped bilayer graphene or highly oriented pyrolytic graphite (HOPG) \cite{Chen2015-nature-commu,Fang-prb2017,2DMaterials-2018-substrate-dependent-TC,Sugawara-ACSNANO-2016} \tcdw$=200-230$K. This strong variability 
of \tcdw\ has been tentatively ascribed to the different substrate dielectric constants in possible relation with an excitonic insulator picture \cite{2DMaterials-2018-substrate-dependent-TC}. 
Indeed, as the exciton binding energy increases in low dimension \cite{KohnGreyTin,JeromePhysRev.158.462,KeldyshKozlov,exciton-insulator-1965,exciton-insulator-PhysRev-1967}, 
conventional wisdom suggests that excitonic insulators should be more stable in 2D than in 3D.
However, other effects such as charge transfer from the substrate, the non-stoichiometry due to Se vacancies or doping could be very relevant. From theory, on the one hand the \tcdw\ of \tise\ monolayer has up to now only been estimated from a variation of the electronic temperature \te . At the harmonic level this assumption predicts a \tcdw$\approx 1195$K within PBE and \tcdw$\approx 1920$K by including the exchange interaction via HSE06 \cite{suppmat}, 
in complete disagreement with the experimental data and leading to an incorrect estimation of \tcdw\ of
at least a factor of $5$. On the other hand, little is known about the effects of electron-hole exchange interaction 
on the vibrational properties of single-layer \tise\ and its dependence on doping.  
It has been shown that, neglecting the spin-orbit coupling, semi-local functionals are successful in reproducing the semiconducting state of the \lowtt\ phase \cite{Guster_2018}, contrary to what happens in the bulk case. 

In this work we study the anharmonic phonon spectra of an isolated single-layer \tise\ within the stochastic self-consistent harmonic approximation (SSCHA) \cite{SSCHA-Ion-prl2013, Raffaello-PRB-2017, Ion-PRB-2014,Monacelli-prb-2018} that has been successfully applied to study the anharmonicity of other transition metal dichalcogenides \cite{Unai-PRL2019,Raffaello-nanoletter2019}. In particular, by including the exchange interaction via semi-local and hybrid functionals \cite{methods}, we determine the CDW transition
and demonstrate that its melting is determined
by the combined effect of phonon-phonon scattering and electron doping and not by an excitonic mechanism.

\begin{figure*}[tb]
	\centering
	\vspace{0.2cm}
	\includegraphics[width=0.31\linewidth]{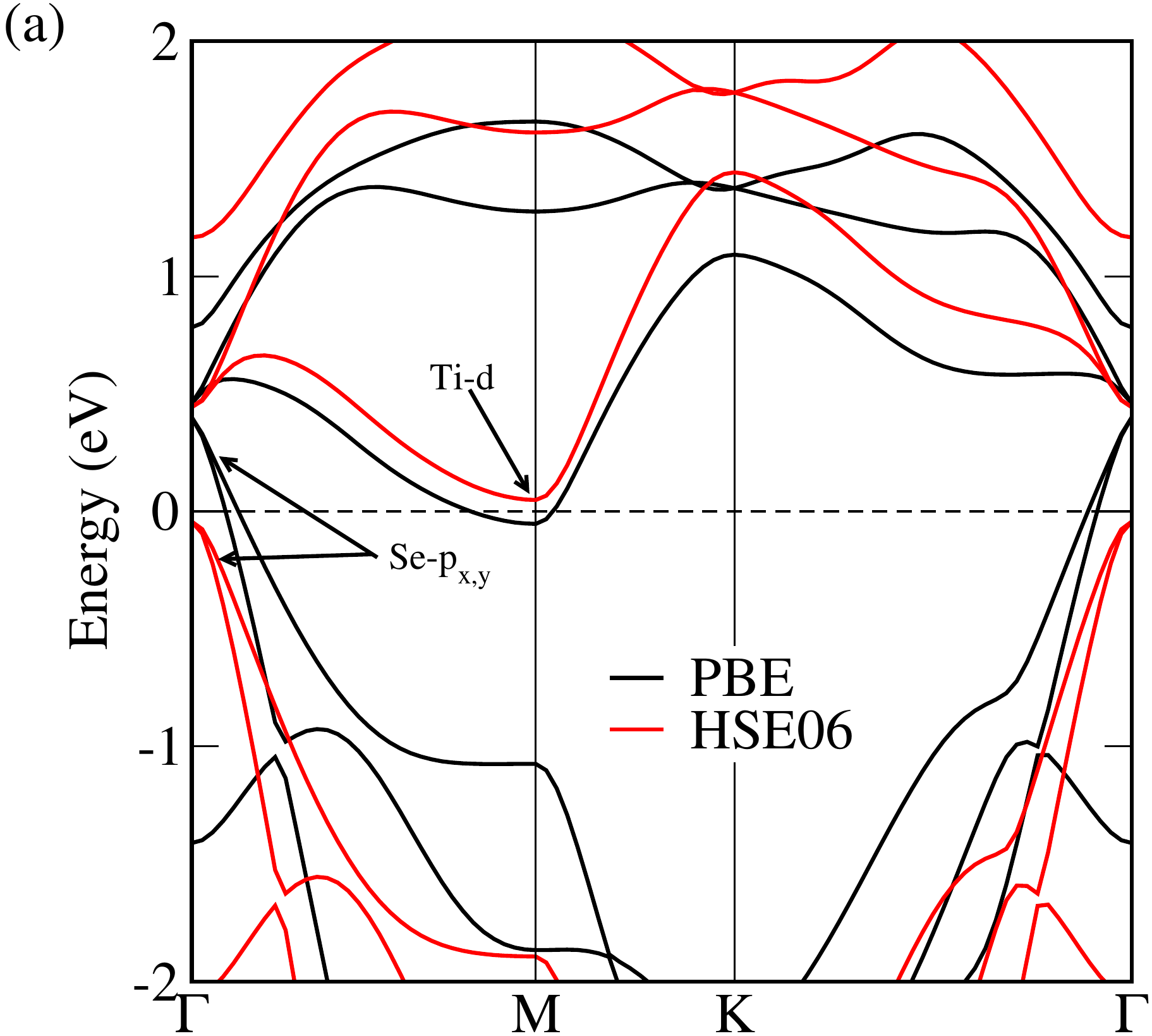} \hspace{0.25cm}
	\includegraphics[width=0.31\linewidth]{harmonic_sc441_pbe_hse06.eps} \hspace{0.25cm}
    \includegraphics[width=0.31\linewidth]{D4V_all.eps} 
    \caption{(a) Electronic bands from PBE and HSE06 approximations. (b) The harmonic phonon bands calculated on a $4\times4$ supercell and the energy gain as a function of $\delta$Ti where $\Delta E$ is the energy difference per $2\times2$ cell. (c) The anharmonic phonon dispersion on a $4\times4$ supercell at 0 and 100 Kelvin from PBE and HSE06 force engines. PBE and HES06 results are obtained from \qe\ and \crystal , respectively.}
    \label{fig:figure-1-compare-pbe-hse06}
\end{figure*}

ARPES measurements show that the \hightt\ phase of single-layer \tise\ is a weakly $n$-doped semiconductor \cite{Chen2015-nature-commu,Sugawara-ACSNANO-2016,PChen-Nanoletters-2016} with a $0.098$ eV indirect band gap between the $\Gamma$ (i.e., derived from Se $4p_{xy}$ states) and M (i.e., derived from Ti $3d$ states) points in the Brillion zone (BZ). In Fig. \ref{fig:figure-1-compare-pbe-hse06} (a) the calculated electronic band structures in both PBE and HSE06 approximations are presented. PBE predicts a metallic state with a negative band gap of $-0.45$ eV between the $\Gamma$ and M points in BZ, in good agreement with previous calculations \cite{Singh-prb2017,PasquierPhysRevB.98.235106,Guster_2018} but in disagreement with ARPES. The HSE06 yields a positive band gap of 0.092 eV, in perfect agreement with ARPES. 

The PBE and HSE06 harmonic phonon dispersion are very similar, despite a different electronic structure, as shown in Fig. \ref{fig:figure-1-compare-pbe-hse06} (b). The A$_{1u}$ mode
at the M point is strongly unstable (imaginary phonon frequencies are represented as negative values in all dispersion plots), indicating formation of a $2\times2$ superstructure. 
The two functionals lead to $\approx 20$ cm$^{-1}$ (i.e., $12\%$) difference in the A$_{1u}$ imaginary phonon frequency.
Other modes at higher energy suffer of a somewhat stronger renormalization. In order to understand if the effect of exchange on the 
CDW is small, we also calculate the energy gain with respect to the displacement of Ti atoms corresponding to the CDW pattern (see Fig. \ref{fig:displacement}) using the experimental ratio $\delta$Ti/$\delta$Se = 3   \cite{Salvo-prb-1976-energy-gain,Fang-prb2017}, as shown in the inset of Fig. \ref{fig:figure-1-compare-pbe-hse06} (b). Again, the two approximations yield a similar minimum indicating a similar CDW instability, which eventually confirms the negligible effect of electron-hole exchange in the soft-mode harmonic phonon bands. Interestingly, also the position of the energy minimum versus $\delta$Ti is practically the same, indicative of an akin CDW structure. Note that in bulk \tise , the energy gain for a distortion having modulation ${\bf q}={\bf \Gamma L}$ in HSE06 is approximately three times larger than the PBE one \cite{MCalandra-PRL2017} and the minimum
occurs at substantially larger $\delta$Ti in HSE06 than in PBE.

This puzzling difference between bulk and single layer can be understood by noting that in the former
the strong  electron-hole interaction is between the 4$p_{xy}$ occupied states at zone
center and some of the empty Ti 3$d$ states  at the L point. In the bulk, for a distortion having periodicity ${\bf q}={\bf \Gamma M}$ (i.e., all the TiSe$_2$ layers distort in phase), and coupling the Brillouin zone regions around the A point with those around the L point, the
 energy gain by the distortion is reversed with respect to the case of a distortion
 having modulation ${\bf q}={\bf \Gamma L}$, namely
 the PBE energy gain is much larger than the HSE06 one \cite{suppmat}. Thus, in the bulk, the exchange  interaction effects depend crucially  on the modulation of the distortion and on the electronic states  involved.
 In the single layer, the  electronic structure of the \hightt\ phase along $\Gamma$M is very similar to the one of the undistorted bulk along the A-L line, with $\Gamma$ and M in the single layer corresponding A and L in the bulk. For this reason, the effect of the  exchange interaction on the phonon dispersion is much weaker for a ${\bf q}={\bf \Gamma M}$ modulation in the single layer than for the case of a ${\bf q}={\bf \Gamma L}$ distortion on the bulk.
On top of that, other effects contribute to the different energy gain by the ${\bf q}={\bf \Gamma M}$ distortion in bulk and single layer, such as the weak but non-negligible band dispersion along k$_z$ close to the A and L high symmetry points in the bulk and the slightly different fillings of the Ti $d$-band at L in the bulk and M in the single layer.
This explains why in single layer the effects of exchange on the charge density wave distortion are negligible and demonstrates how simple arguments based on isotropic coulomb interactions  \cite{KohnGreyTin,JeromePhysRev.158.462,KeldyshKozlov,exciton-insulator-1965,exciton-insulator-PhysRev-1967}
do not apply easily in layered materials
with weak interlayer binding, such as TiSe$_2$. 

Even if the two functionals give practically identical energy versus displacement profiles, this is not enough to conclude that the exchange interaction is irrelevant for the soft mode at the anharmonic level. For this purpose, we calculate the phonon dispersion including nonperturbative anharmonic effects within the SSCHA using both HSE06 and PBE as force engines. Namely, we evaluate the temperature dependent dynamical matrix 
$\bDcal =  \bold{M}^{-\frac{1}{2}}\eval{\pdv{F}{\bRcal}{\bRcal}}_{\bRcal_{eq}} \bold{M}^{-\frac{1}{2}}$ where $\bold{M}$ is the
matrix of the ionic masses $M_a$ with $M_{ab}=\delta_{ab}M_a$ within the SSCHA. The free energy curvature with respect to the vector of the centroid positions
$\bRcal$ reads \cite{Raffaello-PRB-2017}:
\begin{equation}
\label{eq:free-energy-hessian}
   \pdv{F}{\bRcal}{\bRcal}=\bPhi +\overset{(3)}{\bPhi}\bLambda(0) \overset{(3)}{\bPhi}+\overset{(3)}{\bPhi}\bLambda(0){\bf \Theta} \bLambda(0)\overset{(3)}{\bPhi} \, ,
\end{equation}
where $\bPhi$ represents the SSCHA force constant, $\overset{(3)}{\bPhi}\bLambda(0)\overset{(3)}{\bPhi}$ is the so-called ``static bubble term'', and $\overset{(3)}{\bPhi}\bLambda(0){\bf \Theta} \bLambda(0)\overset{(3)}{\bPhi}$ contains the higher order terms. Here $\overset{(n)}{\bPhi}$ refers to the $n$-th order anharmonic force constants averaged over the density matrix of the SSCHA hamiltonian
(see Ref. \cite{Raffaello-PRB-2017} for more details on notation). All these quantities can indeed be recasted as appropriate stochastic averages over the atomic forces.
As the HSE06 calculation is computationally expensive, we perform the calculation on a $4\times4$ supercell (i.e., 48 atoms).
Even if this supercell size is not completely converged and \tcdw\ is underestimated, as it will be shown later, it is clear from Fig. \ref{fig:figure-1-compare-pbe-hse06} (c) that PBE and HSE06 yield practically the same low energy dispersion around the M point even with full inclusion of anharmonicity. Moreover, the temperature dependence of the soft mode is also very similar, indicative of a practically identical \tcdw\ (i.e., identified as the point where the energy of the soft phonon at M crosses zero) for the two functionals on the $4\times4$ cell. Other phonon modes, particularly around zone center, suffer of a somewhat
stronger renormalization by exchange (analogous to the harmonic case), however, 
as we are mainly interested in the CDW transition, we can stick to the PBE functional and proceed with calculations on larger supercells (see \cite{suppmat} for additional technical details and the
magnitude of the different terms in Eq. \eqref{eq:free-energy-hessian}).

\begin{figure}[tb]
    \centering
    \includegraphics[width=0.9\linewidth]{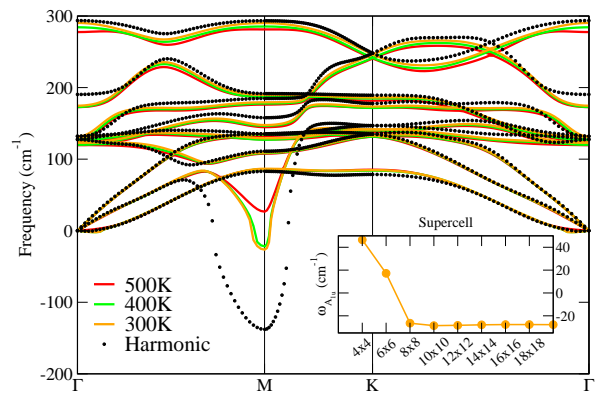}
    \caption{Harmonic and anharmonic phonon dispersion on a $8\times8$ supercell. Inset: the convergence of the lowest phonon frequency at the M point $\omega_{A_{1u}}$ with respect to the size of the cell at 300K. The results of the cell larger than $8\times8$ come from the interpolation method detailed in \cite{suppmat}.}
    \label{fig:figure-2-D3V-881-200-400-425K}
\end{figure}

The anharmonic phonon spectrum obtained by evaluating Eq. \eqref{eq:free-energy-hessian} on a $8\times8$ supercell (i.e., 192 atoms) for several temperatures is shown in Fig. \ref{fig:figure-2-D3V-881-200-400-425K}. 
As it can be seen, the harmonic phonon frequency of the 
lowest energy mode at M is $\omega_{A_{1u}}\approx -135$ cm$^{-1}$, while at 300K the anharmonic phonon frequency of the same mode is $\approx -26$ cm$^{-1}$. Thus, already at room temperature, the anharmonic correction is of the same order of the harmonic phonon frequency.
Between 400 and 500K, this phonon mode becomes positive, compatible with a CDW transition within this temperature range. Note that this transition temperature differs substantially with respect to the one on a $4\times4$ supercell which reflects the importance of cell size. To better illustrate this point, in the inset of  Fig. \ref{fig:figure-2-D3V-881-200-400-425K} we show the convergence of the soft mode phonon frequency as a function of the cell size at 300K. The A$_{1u}$ phonon frequency at M is fully converged on the $8\times8$ cell \cite{suppmat}. We can then obtain \tcdw$\approx440$K for a suspended and undoped monolayer \tise .

\begin{figure}[tb]
    \centering 
    \includegraphics[width=0.9\linewidth]{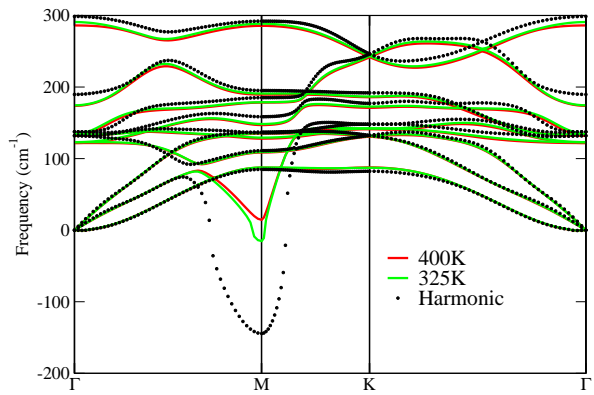}
    \caption{Harmonic and anharmonic phonon spectra for doped ($n=4.6\times 10^{13}$cm$^{-2}$) samples calculated on a $8\times8$ supercell. }
    \label{fig:Tcdwdoping}
\end{figure}

Our calculated \tcdw\ from first-principles SSCHA is 1.6-2.0 times higher than the measured one (depending on the substrate \cite{Sun2017,2DMaterials-2018-substrate-dependent-TC, Wang2018-advmat, Li2016-nature}). However, our calculation is for an undoped isolated monolayer, the measured samples are instead supported by the substrate and substantially $n$-doped. In order to understand the origin of this discrepancy, we investigate the effect of electron-doping to see if it can be responsible for the decrease in \tcdw . To this end, we first determine the electron doping amount by performing HSE06 $n$-doped electronic structure calculations by changing the number of valence electrons and adding a compensating jellium  (i.e., the virtual crystal approximation (VCA)) until the bands agree well with the ARPES spectra \cite{Chen2015-nature-commu} (see \cite{suppmat} for details). With this electron density (i.e., $4.6\times 10^{13}$ cm$^{-2}$ indicative of a substantial doping), we then perform a PBE linear response harmonic calculation to obtain the harmonic phonon dispersion shown in Fig. \ref{fig:Tcdwdoping}. It turns out that the harmonic phonon dispersion (and consequently the effects of the electron-phonon interaction) are doping independent at this level of doping (by comparing the two black dotted curves in Figs. \ref{fig:figure-2-D3V-881-200-400-425K} and \ref{fig:Tcdwdoping}), consistent with an earlier study \cite{Guster_2018}. However, when anharmonicity is included, the electron-doping substantially suppresses the CDW instability as illustrated in Fig. \ref{fig:Tcdwdoping} leading to \tcdw$\approx364$K for a suspended \tise\ monolayer, close to the
experimental data of $280$K on insulating MoS$_2$ substrate \cite{2DMaterials-2018-substrate-dependent-TC}.

In conclusion, we study anharmonic effects in a free-standing \tise\ monolayer within the stochastic self-consistent harmonic approximation. We have shown that the electron-hole exchange plays only a marginal role on the vibrational properties of its \hightt\ phase, at odds with its bulk counterpart where the exchange interaction is crucial. We showed that the weakening of the electron-hole interaction in single layer is related to the different periodicity of the modulation with respect to the bulk and the fact that it couples different states in the electronic structure.
Our results upturns the conventional wisdom that the electron-hole interaction should be stronger in low dimension due to an increase in binding energy 
\cite{KohnGreyTin,JeromePhysRev.158.462,KeldyshKozlov,exciton-insulator-1965,exciton-insulator-PhysRev-1967}, mainly because of the strong momentum dependence of the electron-hole interaction and the complex multiband nature of the electronic structure in TiSe$_2$.  
It also underlines that simple qualitative arguments based on the exciton binding energy and its dependence on the effective mass and on the screening (see Eq. (1) in Ref. \cite{KohnGreyTin})
do not easily apply since they are unable to explain the occurrence of charge density waves and the temperature dependence of phonon spectra when reducing the dimensionality.

By studying the temperature dependence of the A$_{1u}$ soft mode at the M point, we find the \tcdw\ of an isolated and undoped single-layer to be $\approx 440$K,  while \tcdw$\approx 364$K for an electron-doping $n=4.6\times10^{13}$ cm$^{-2}$, close to the experimental value for supported sample. Thus, \tcdw\ is strongly 
doping dependent when including anharmonicity, an effect completely
absent at the harmonic level as harmonic spectra are weakly doping dependent. Our work establishes phonon-phonon scattering and the density of carriers in the conduction band as the two mechanisms determining the melting of CDW in a single-layer \tise.

Computational resources were granted by PRACE (Project No. 2017174186) and from
IDRIS, CINES and TGCC (Grant eDARI 91202 and Grand Challenge Jean Zay). We acknowledge support from the Graphene Flaghisp core 2 (Grant No. 785219) and Agence nationale de la recherche (Grant No. ANR-17-CE24-0030). F. M. and L. M. acknowledge support by the MIUR PRIN-2017 program, project number 2017Z8TS5B. I.E. acknowledges financial support from the Spanish Ministry of Economy and Competitiveness (FIS2016-76617-P). We acknowledge U. Aseguinolaza for useful discussions.

\bibliographystyle{apsrev4-1} 
\bibliography{main}
\end{document}